\documentclass[%
 reprint,
superscriptaddress,
 amsmath,amssymb,
 aps,
pra,
floatfix,
twocolumn,
]{revtex4-2}
\usepackage{graphicx}
\usepackage{natbib}
\usepackage{dcolumn}
\usepackage{bm}
\newcommand{\pd}[2]{\frac{\partial #1}{\partial #2}}
\usepackage{physics}
\usepackage{isomath}
\usepackage{booktabs}
\usepackage{soul, xcolor}
\usepackage{placeins}
\usepackage{float}

\begin{document}


\title{Reinforcement Learning for Rotation Sensing with Ultracold Atoms\\ in an Optical Lattice}

\author{Liang-Ying Chih}
\affiliation{JILA and Department of Physics, University of Colorado, 440 UCB, Boulder, CO 80309, USA}

\author{Murray Holland}
\email[]{murray.holland@colorado.edu}
\affiliation{JILA and Department of Physics, University of Colorado, 440 UCB, Boulder, CO 80309, USA}
\date{\today}

\begin{abstract}
In this paper, we investigate a design approach of reinforcement learning to engineer a gyroscope in an optical lattice for the inertial sensing of rotations. Our methodology is not based on traditional atom interferometry, that is, splitting, reflecting, and recombining wavefunction components. Instead, the learning agent is assigned the task of generating lattice shaking sequences that optimize the sensitivity of the gyroscope to rotational signals in an end-to-end design philosophy. What results is an interference device that is completely distinct from the familiar Mach-Zehnder-type interferometer. For the same total interrogation time, the end-to-end design leads to a 20-fold improvement in sensitivity over traditional Bragg interferometry.
\end{abstract}

\maketitle

Over the past decade, optical Michelson interferometers have achieved astounding levels of sensitivity leading to entirely new fields of gravitational wave cosmology~\cite{LIGO,VIRGO}.
Their matter-wave counterparts, in particular atom interferometers~\cite{atomInterferometry1, atomInterferometry2}, have proven to be a principal technology for sensing~\cite{PhysRevLett.111.083001, tractorGyro, Multi_loop_gyro, boshier2022, PhysRevLett.129.183202, Cooling_2021, panda2022quantum, panda2023probing, TwinLattice, BlochOscillations, PointSource, AtomInterferometry3}, especially for the inertial measurements of accelerations, rotations, gravimetry~\cite{Peters1999}, and gravity gradiometry~\cite{PhysRevLett.81.971}. The sensitivity that can be achieved in atom interferometry is fundamentally limited by the de Broglie wavelength of the atoms, the area that can be enclosed in space and/or space-time, and by the number of particles~\cite{Pippa}. Improvements have primarily focused on individual components, such as larger momentum splitting and longer hold time. 

Recently, however, a completely distinct approach to light pulse atom interferometry has been developed~\cite{Weidner1,Chih}. It involves confining atoms to an optical lattice during the entire interferometry protocol, and builds on early experiments that created beamsplitters and other components by shaking the lattice~\cite{JHeckerDenschlag_2002}. In part, the purpose of the optical lattice is to provide robustness of the system in the face of a harsh dynamical environment typical of real-world applications~\cite{Cavity_Interferometer, Guided_Interferometer}, including platform vibrations and temperature fluctuations and drifts. Since there is no unique protocol for generating the components of beam-splitters and mirrors in an optical lattice, it makes sense to delegate this optimization task to machine learning algorithms~\cite{Chih}. One `teaches' the system to carry out an interferometric task by learning how to correctly modulate the phase of the lattice. Experiments using this method can now demonstrate full interferometer sequences that cascade the machine-designed components together~\cite{ledesma2023machinedesignedopticallatticeatom,ledesma2024vectoratomaccelerometryoptical,ledesma2024universalgatesetoptical}. In these experiments it has been shown that the learning can be performed in simulation first and only later applied to the experiment, as demonstrated by the high-fidelity agreement of the measured data with the anticipated sensor performance.

\begin{figure}
\centering
\includegraphics[width=0.4\textwidth]{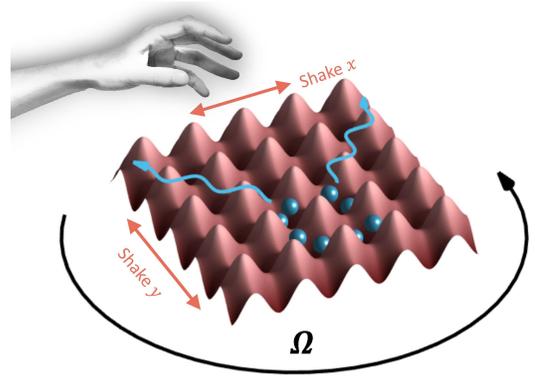}
\caption{A two-dimensional optical lattice potential in a rotating frame, where atoms are trapped and their dynamics can be controlled by an external agent (the hand) shaking the lattice. The blue arrows indicate possible trajectories associated with splitting of the matterwaves. }
\label{fig:lattice_3d}
\end{figure}

While these works have established the key principles, the learning that was implemented was constrained by a premeditated viewpoint as to how interferometry is to take place. Traditionally, nearly every kind of interferometry, optical or matter wave, results from a sequence of wavefront splitting, reflection, and recombination, with wave propagation in between, known as the Mach-Zehnder configuration \cite{Mach,Zehnder}. However, if the learning agent is freed from being restricted by conventional wisdom, it could potentially discover solutions that surpass what humans have explored to date~\cite{PhysRevA.98.023629}. This different learning goal may provide a pathway to the discovery of a revolutionary kind of atomic gyroscope that is purely optimized for specified design constraints, such as a combination of sensitivity, dynamic range, and tolerance to noise and experimental drifts.

A type of machine learning that has the potential to realize this goal is reinforcement learning \cite{RLSutton}. Reinforcement learning is especially effective when the agent is trained in an end-to-end manner, meaning an uninterrupted shaking sequence from start to finish that is not decomposable into components such as beam splitters and mirrors. This is because reinforcement learning allows for the maximization of long-term rewards of a control protocol in situations where the optimal solution is not transparent at each individual step. Reinforcement learning has emerged as a powerful tool in quantum research and has already been applied to a variety of quantum problems, such as the generation of highly-entangled states and the development of quantum error correction protocols, thereby bringing new insights to the field \cite{ActiveLearning, RL_qSensor, RL_Bell, RL_KapitzaOscillator, RL_manyBody, RL_qCommunication, RL_qCircuit, RL_QEC1, RL_QEC2, RL_cartpoles, RL_QPE, RL_statePrep, RL_entanglement}.

In this paper, we present results from the application of reinforcement learning to quantum metrology, specifically to train a two-dimensional optical lattice to sense rotation with high precision and thereby embody the design objectives of a gyroscope. The system we consider is one where the atoms are confined in a two-dimensional optical lattice, as shown in Fig.~\ref{fig:lattice_3d}. 

\begin{figure}[b]
\begin{center}
\includegraphics[width=0.5\textwidth]{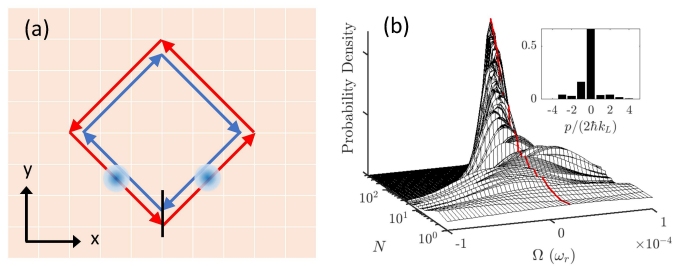}
\end{center}
\caption{(a) The clockwise trajectory (blue) and the counter-clockwise trajectory (red) taken by the wave functions in the two-path shaken lattice gyroscope. The vertical back line is where the wave functions were initially split and are eventually recombined. (b) Bayesian reconstruction of the rotation rate from the interference patterns generated by the two-path gyroscope. The main plot shows the probability distribution of the estimated rotation rate for numbers of measurements ranging from 1 to 500 and lattice depths $V_x=V_y=10 \hbar\omega_r$, where $\omega_r=\hbar k^2/(2m)$. The red curve indicates the true value of the rotation rate, which was $2\times 10^{-4}~\omega_r$. At 500 measurements, the peak of the distribution coincides with the true value, meaning that the estimation is unbiased. The inset shows the momentum distribution that results at the output for a rotation rate of $2\times 10^{-4}~\omega_r$.}
\label{fig:schematics}
\end{figure}

The lattice in each dimension can be `shaken' by varying the phases of the corresponding pairs of interfering laser beams.
The entire system is described mathematically in a rotating non-inertial frame by the Hamiltonian
\begin{equation}
    \hat{H} = \frac{\hat{\vectorsym{p}}^2}{2m} - \vectorsym{\Omega}\cdot\hat{\vectorsym{L}} - \sum_{\alpha\in\{x,y\}}\frac{V_{\alpha}}{2}\cos(2k\hat{\alpha} + \phi_{\alpha})\,,
    \label{eq:Hamiltonian}
\end{equation}
where $m$ is the mass of a single atom with coordinate  $\hat{\vectorsym{r}}=\{\hat{x}, \hat{y}\}$ and momentum $\hat{\vectorsym{p}}$, $k$ is the laser wavenumber, $\phi_{\alpha}$'s are the phase differences between the two counter-propagating lasers in the $x$ or $y$ directions, and $V_{\alpha}$'s are the corresponding lattice potential strengths. The term~$\vectorsym{\Omega}\cdot\hat{\vectorsym{L}}$ describes the rotational kinetic energy of the system, with $\hat{\vectorsym{L}}=\hat{\vectorsym{r}}\times\hat{\vectorsym{p}}$ the angular momentum, and~$\vectorsym{\Omega}$  the angular velocity. It is the unknown magnitude, $\Omega$, that is the metrological parameter that we wish to measure with high accuracy. For simplicity, we limit the discussion here to the case of dilute atom clouds, meaning atom-atom interactions do not adversely affect the detection of the rotation signal.

Before investigating end-to-end solutions, we first derive a conventional two-path Sagnac matter-wave interferometer, motivated by the fiberoptic gyroscope \cite{FOG}, but implemented with atoms~\cite{Gyroscope1,Gyroscope2,boshier2022}. This can be constructed in the lattice using the previously developed beam-splitting and reflecting protocols that were based on components optimized by reinforcement learning~\cite{Chih}. The splitting protocol was demonstrated to transfer the ground state to an approximate superposition of the $\ket{\pm 4\hbar k}$ states, and the reflecting protocol to map any linear combination of the $\ket{\pm 4\hbar k}$ states to the corresponding combination of the $\ket{\mp 4\hbar k}$ states. 
In order to build a gyroscope from these components, as illustrated in Fig.~\ref{fig:schematics}(a), the sequence of operations is the following:
\begin{itemize}
    \item The wavefunction is prepared in the ground state of the 2D lattice. 
    \item In the $x$-direction, the 1D beam-splitting protocol is applied, allowing free propagation for a duration, denoted as $T$, and then the reflecting protocol is applied. Following this, free propagation occurs for a duration~$2T$, and then the reflecting protocol is applied again, followed by free propagation for a duration~$T$, and finally the recombining protocol. 
    \item In the $y$-direction, the lattice operates as a conveyor belt and is smoothly accelerated to the velocity $4\hbar k/m$ according to the adiabaticity criterion, as described in the Supplemental Material (SM), and then translated at constant velocity. During the halfway point of the sequence, when the two paths in the $x$-directions cross, the velocity of the $y$-lattice is decelerated through zero to $-4\hbar k/m$, so that the lattice can be translated backward at the constant velocity. The last step is to adiabatically accelerate the $y$-lattice back to zero velocity for the final recombination. 

\end{itemize}
Such a device operates on the principle of the Sagnac effect, which describes the phase difference, $\Delta \phi$, that accumulates between two waves that propagate in  opposite directions in a loop when in a non-inertial rotating frame. In general, the phase difference is proportional to both the angular velocity and to the area that the two paths enclose,~$\vectorsym{A}$. In the case of matter-waves, $\Delta \phi = 4m\vectorsym{\Omega}\cdot \vectorsym{A}/\hbar$~\cite{Sagnac1,Sagnac2}. If higher sensitivity is desired, one could effectively increase $\vectorsym{A}$ by completing many cycles before applying the last recombination step. In practice, one must balance multiple cycles against any reduced fringe visibility arising from imperfections.

With respect to this reference system, we now consider the possibility for a completely end-to-end design that is not based on the decomposition into components. We implement the machine learning for this overarching design goal by employing the Double-Deep-Q-Network approach \cite{doubleDQN}, an algorithm that is particularly effective in situations where deterministic outcomes are expected and sample efficiency is critical (details described in Ref.~\cite{Chih}). This reinforcement learning algorithm consists of an agent that selects actions based on the observed state of the environment, and an environment that carries out the the action and generates a consequential reward based on the resulting quality of the state. The agent does not need to be exposed to the full quantum state, but only to relevant features of that state, such as the population in the discrete momentum basis in $x$ and the average position and momentum in $y$. 
    
\begin{figure}
\centering
\includegraphics[width=0.5\textwidth]{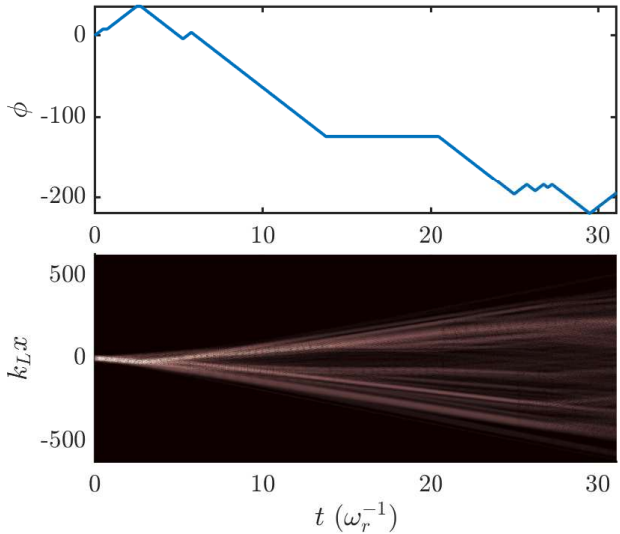}
\caption{Evolution of the phase function and the spatial wavefunction. The shaking protocol was generated from the reinforcement learning agent with $\partial\phi/\partial t$ as the actions in an end-to-end design. There are 125 discrete steps in this protocol. The wavefunction was initialized in the ground state of the lattice convolved with a momentum width $\sigma_p=0.1\,\hbar k$, and then evolved according to the shaking protocol given.}
\label{fig:RL_results_1}
\end{figure}

Motivated in part by potential experimental implementation, we choose the actions to be selected from a finite set of discrete options for the time derivative of the phase,~$d\phi/dt$. This represents a frequency difference between interfering laser beam pairs. The wave packet evolution that this generates is calculated by numerically solving the time-dependent Schr\"odinger equation, using a separation ansatz for the $x$ and $y$ dimensions (see SM). In order to compute the reward, we use the classical Fisher information \cite{Fisher_1922}, which quantitatively measures the sensitivity. To do this, we replicate the environment as three copies of the quantum system with slightly different rotation rates, $\Omega = [\Omega_0-d\Omega,\, \Omega_0,\, \Omega_0+d\Omega]$, for small~$d\Omega$ and with each evolving according to Eq.~(\ref{eq:Hamiltonian}). This construction allows us to compute the derivative with respect to~$\Omega$ by numerical symmetric finite differencing. The classical Fisher information is given by
\begin{equation}
    I(\Omega) = \sum_p \frac{1}{\mathrm{Pr}(p|\Omega)}\left[\pd{\mathrm{Pr}(p|\Omega)}{\Omega}\right]^2 \,,
\end{equation}
where $\mathrm{Pr}(p|\Omega)$ is the probability for measuring a momentum $p$ at the gyroscope output for the given $\Omega$. Reinforcement learning aims to generate steps that maximize the long-term reward, that is, to find a sequence of $d\phi/dt$ steps that optimizes the Fisher information evaluated at the terminal time. The potential sensitivity is constrained by a theoretical bound (Cram\'er-Rao bound) for the standard deviation of the measurement,
\begin{equation}
    \sigma_\Omega \geq 1/\sqrt{NI(\Omega)},
\end{equation}
where $N$ is the number of independent measurements (atoms). In this work, we have optimized for a specific choice of the angular frequency, $\Omega_0$, but straightforward extensions are possible, including designing for net performance over a finite domain.

\begin{figure}
\centering
\includegraphics[width=0.5\textwidth]{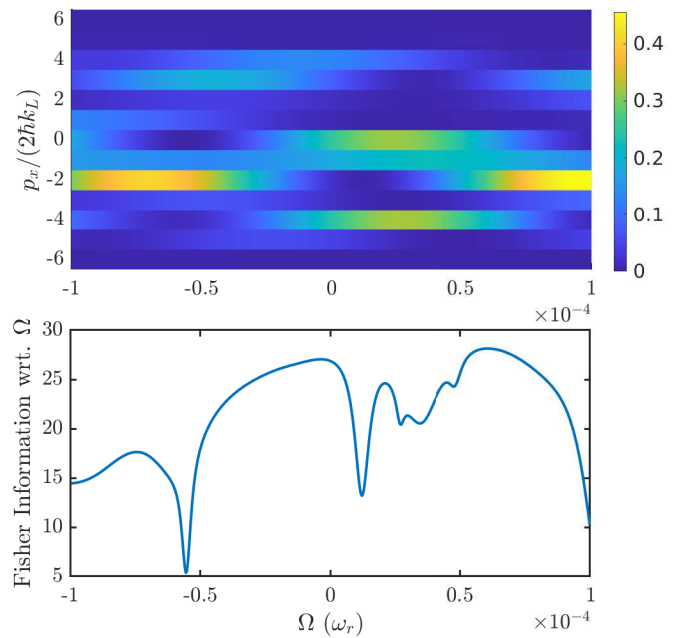}
\caption{Population distribution in the momentum basis and the resulting classical Fisher information for the machine-learned shaken-lattice gyroscope as a function of the rotation rate $\Omega$.   The value of the Fisher information is normalized by the Fisher information for the two-path interferometer, which has splitting of order $\Delta p = 8\hbar k$, and thus sensitivity about 4 times better than a conventional Bragg interferometer.}
\label{fig:RL_results_2}
\end{figure}

Within the context of our learning framework, we are able to obtain lattice shaking protocols that outperform the conventional two-path gyroscope. 
The solution varies for every instantiation due to the randomized initialization of the agent. The illustration in Fig.~\ref{fig:RL_results_1} shows one of the realized machine-learned protocols that gives high sensitivity. The pattern is reminiscent of speckle patterns that emerge from a multi-mode fiber (fiber specklegram), which are known to be sensitive interference detectors of inertial phase. However, in the case of a multi-mode fiber system, the patterns are not robust and are scrambled by temperature changes or strain on the fiber. The situation is quite different here, since the adverse noise and imperfections that may enter are different in origin and primarily common-mode. Despite the non-trivial and irregular pattern, this device will measure rotation signals with high accuracy, something that we now demonstrate by simulating an example experimental measurement.

\begin{figure}[b]
\centering
\includegraphics[width=0.4\textwidth]{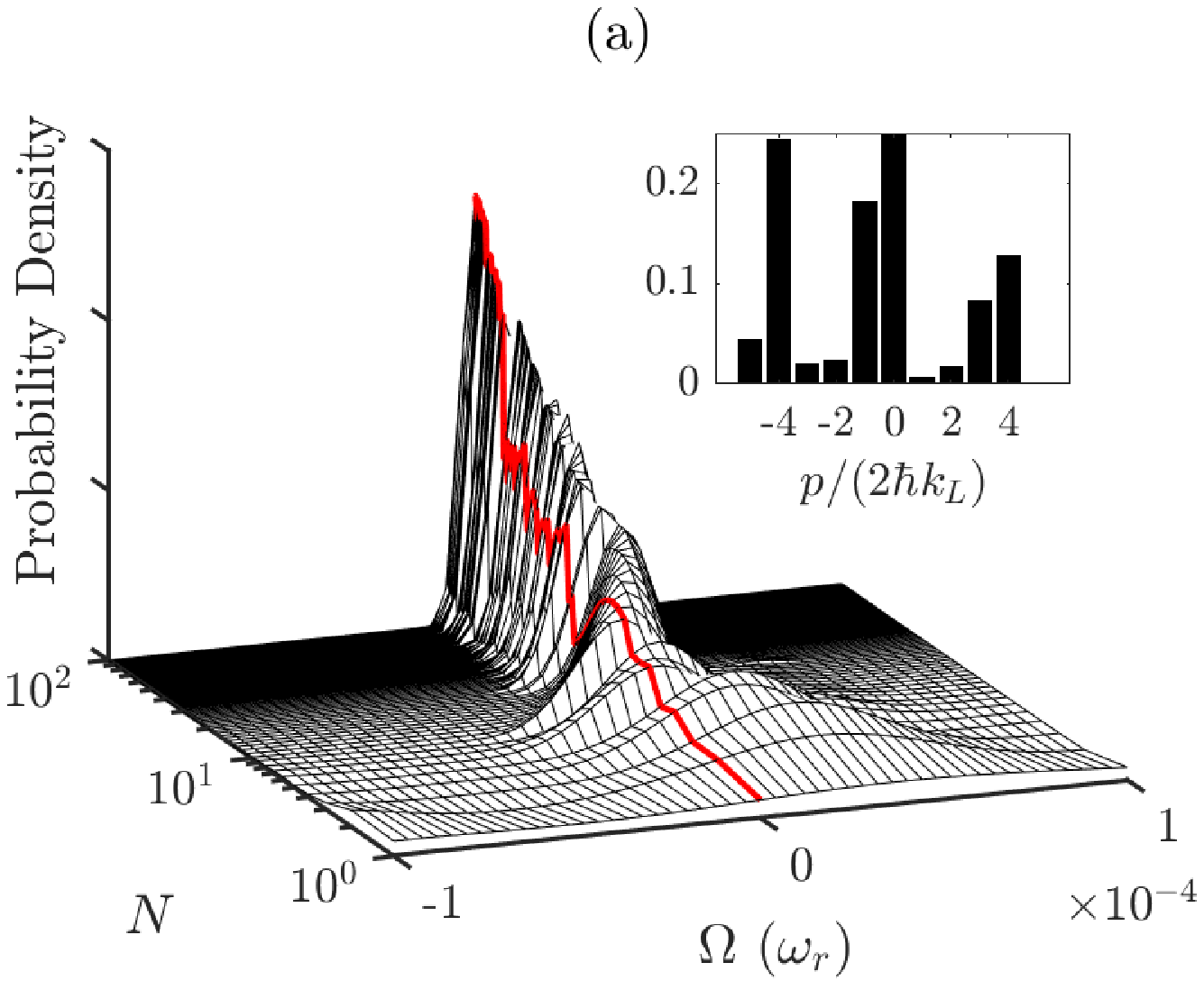}
\includegraphics[width=0.4\textwidth]{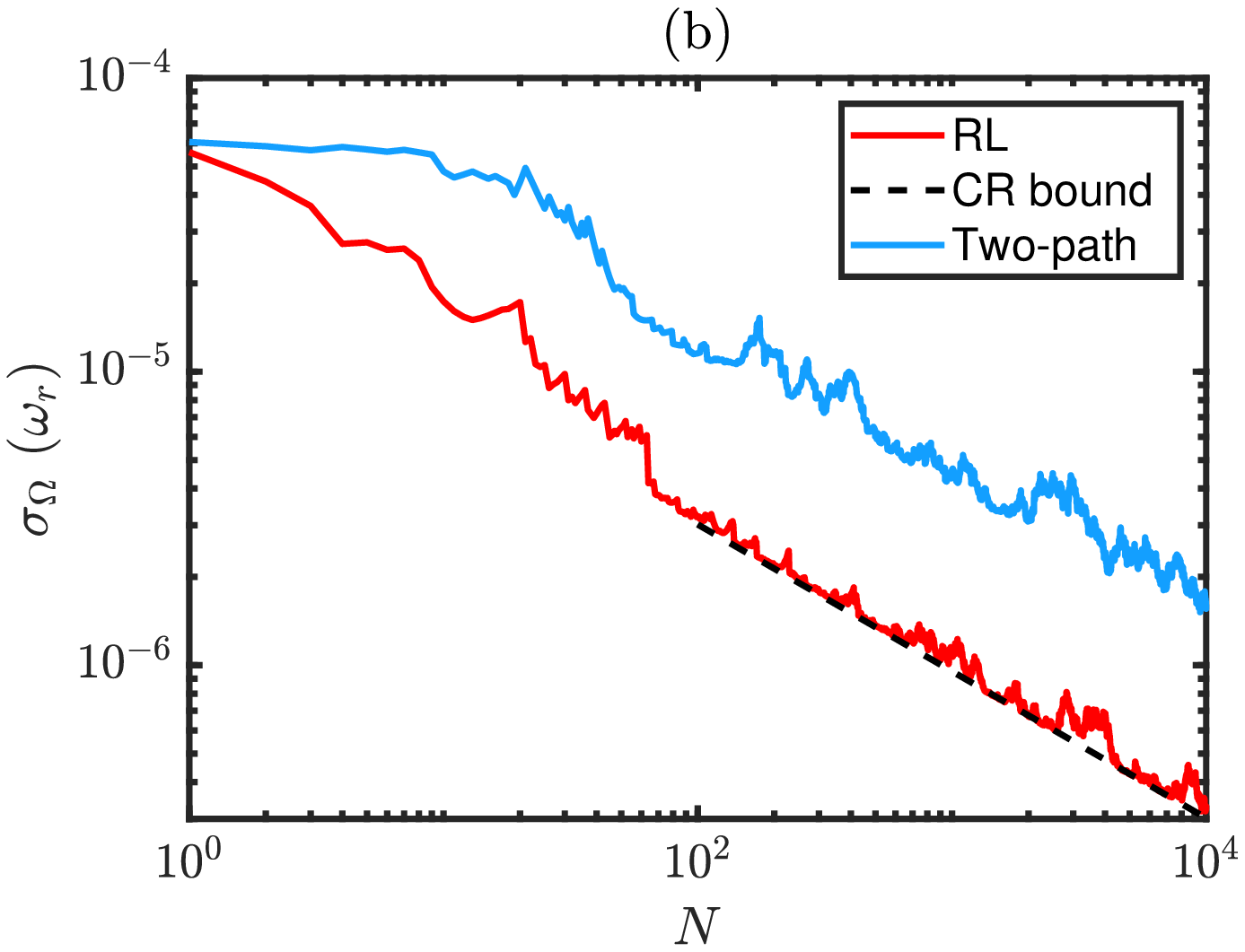}
\caption{(a) An example Bayesian reconstruction of the rotation rate, showing the probability distribution for sampling 1 atom to 100 atoms. The true rotation value is indicated by the red line, and the inset shows the shaken lattice interferometer output distribution for this value. (b) The standard deviation of the rotation rate as a function of the atom number (RL, red curve). Note that the behavior of the curve is subject to large sampling noise at small $N$, and only begins to agree with its asymptotic limit at around $N=100$. The result at large atom number, when the central limit theorem applies and the sample average converges to the true value, is consistent with the Cram\'er-Rao (CR, dotted line) bound. The performance greatly surpasses the two-path interferometer shown earlier (Two-path, blue curve).}
\label{fig:fisher} 
\end{figure}

The momentum distributions that are produced by the multi-path interferometer under conditions of different values of the rotation rate are shown in Fig.~\ref{fig:RL_results_2}. The population of each momentum component is indicated by false color. Momentum distributions are typically directly observable in experiment by expanding the cloud under time-of-flight, and integrating regions of the absorption image to find the relative proportions for each momentum component. The more detailed structure there is in the observed momentum distribution as the rotation rate is varied, the more sensitive is the interferometer.
Note that each vertical slice is essentially unique, and therefore the momentum distribution is a fingerprint that allows one to infer the rotation rate without aliasing. 
The Fisher information calculated from the distribution is also shown in Fig.~\ref{fig:RL_results_2}.
The value on the $y$-axis is the ratio between the Fisher information from this multi-mode interferometer and the one shown earlier based on the two-path arrangement. We optimize the Fisher information around $\Omega_0=0$, and one can see that the Fisher information is maximum at $\Omega=0$, as anticipated, with large dynamic range being sacrificed for high sensitivity.

The result of our design was a machine-learned gyroscope that achieved a Fisher information of around 25 times higher than the previously investigated two-path gyroscope, meaning that the sensitivity is improved by the square root of this, or a factor of~5. In other words, the interferometer is effectively as sensitive as a conventional one that is 5 times larger. This gain can be understood from several aspects, including population transfer into the higher momentum states (the maximum observed splitting is of order $18\hbar k)$, and a larger spatial footprint, as also seen in Ref.~\cite{PhysRevA.98.023629}. If we make a comparison to the gyroscope with the same total interrogation time as constructed from conventional Bragg interferometry, the improvement is $4\times5=20$-fold.
The extra factor of 4 introduced here is primarily due to large angle splitting of the two-path gyroscope with respect to the Bragg one, i.e., $8\hbar k$ instead of $2\hbar k$, as was shown in Ref.~\cite{Chih}. The  potential gain of 20 is remarkable and reveals the scope for significant improvements upon current state-of-the-art devices.

To extract the rotational signal from the distribution shown in Fig.~\ref{fig:RL_results_2}, we apply Bayesian reconstruction \cite{Bayesian}. This means that we iteratively update the prior distribution for $\Omega$ from each atom measurement. The reconstruction of the rotational signal is shown in Fig.~\ref{fig:fisher}(a), and note that the peak coincides with the true rotation rate as indicated by the red line, thereby verifying that the estimation for~$\Omega$ is unbiased. In Fig.~\ref{fig:fisher}(b), we plot the standard deviation of $\Omega$, and see that the sensitivity scales inversely with the square root of the number of atoms for up to $10^4$ atoms, as expected from independent measurements. We emphasize that for large atom number, interactions between atoms may become increasingly important and are not considered here.

In conclusion, we have shown how to build a gyroscope in an optical lattice that uses demonstrated experimental methodology~\cite{ledesma2023machinedesignedopticallatticeatom}, but extended here to end-to-end design.
From a broader perspective, these results show the exceptional potential of the general learning approach for quantum design, and opens up the application of the same principal machinery to a variety of quantum sensing problems with more complex landscapes, such as multi-parameter estimation and entanglement-enhanced metrology \cite{NonlinearInterferometer,reilly2023optimal}. Furthermore, the implications of our results may go beyond this framework, and the same ideas of reinforcement learning may be applied to a variety of quantum and classical systems, such as those where the design of complex quantum circuits for algorithmic tasks is needed~\cite{PhysRevLett.123.260505}.

The authors acknowledge helpful discussions with Dana Anderson, Marco Nicotra, Penina Axelrad, Catie LeDesma and Kendall Mehling. This work was supported by NSF OMA 1936303,
NSF PHY 2207963, NSF OMA 2016244, and NSF PHY 1734006.

%

\end{document}